\documentstyle[sprocl,epsf]{article}

\newcommand{\be}{\begin{equation}}
\newcommand{\ee}{\end{equation}}
\newcommand{\etal}{{\it et al.}}
\newcommand{\hmp}{h^{-1}Mpc}
\newcommand{\bef}{\begin{figure}}
\newcommand{\eef}{\end{figure}}
 
\def\spose#1{\hbox to 0pt{#1\hss}}
\def\ltapprox{\mathrel{\spose{\lower 3pt\hbox{$\mathchar"218$}}
 \raise 2.0pt\hbox{$\mathchar"13C$}}}
\def\gtapprox{\mathrel{\spose{\lower 3pt\hbox{$\mathchar"218$}}
 \raise 2.0pt\hbox{$\mathchar"13E$}}}
\def\inapprox{\mathrel{\spose{\lower 3pt\hbox{$\mathchar"218$}}
 \raise 2.0pt\hbox{$\mathchar"232$}}}
 
\def\be{\begin{equation}}
\def\ee{\end{equation}}
\def\bea{\begin{eqnarray}}
\def\eea{\end{eqnarray}}

\begin{document}

\title{Fractal structures and the large scale distribution of galaxies
\footnote{In the proceedings of the  
7th Course in astrofundamental physics, Nato Advanced Study Institute, 
International 
Euroconference Erice,  5-16 December 1999}}
\author{Luciano Pietronero$^1$ and Francesco Sylos Labini$^{2,1}$} 
 
\address{      	 
		$^1$INFM Sezione Roma1,         
		Dip. di Fisica, Universit\'a "La Sapienza",  
		P.le A. Moro, 2,   
        	I-00185 Roma, Italy.  
        	\\ 
        	$^2$D\'ept.~de Physique Th\'eorique,  
		Universit\'e de Gen\`eve,   
		24, Quai E. Ansermet, CH-1211 Gen\`eve, Switzerland. 
		}

\maketitle

\abstracts{Galaxy structures are certainly fractal
up to a certain crossover scale $\lambda_0$.
A clear determination of such a scale is still missing.
Usually the conceptual and practical implications of this
property are neglected and the structures are only discussed in 
terms of their global amplitude. 
Here we present a compact summary of these implications.
First, we discuss the problem of the identification of the crossover
scale $\lambda_0$ and the proper characterization
of the scaling. We then consider the implications
of these properties with respect to various physical phenomena
and to the corresponding characteristic values,
i.e.  $r_0$, $\sigma_8$, $\Omega$, etc.  These implications
crucially depend on the value of $\lambda_0$, but they are
still important for a relatively small value, say
$\lambda_0 \approx 50 \hmp$.
Finally we consider the main theoretical
consequences of these results.
}
 
\section{Introduction}

Nowadays there is a general agreement about the fact 
that galactic structures are fractal up to a distance 
scale of $\sim 50 \hmp$ \cite{slmp98,jmsl99} and the  
increasing interest about the fractal versus  
homogeneous distribution of galaxy in the last year 
 \cite{coles98,scara98,rees99,cappi98,martinez99,hutton99,chown99,landy99,nl00} 
has focused, mainly on the determination of the homogeneity 
scale $\lambda_0$.\footnote{See the web page {\it  
http://pil.phys.uniroma1.it/debate.html } where  
all these materials have been collected}
The main point in this discussion 
is that galaxy structures are fractal no matter what is
the crossover scale, and this fact has never been
properly appreciated. Clearly, qualitatively different
implications are related to different values
of $\lambda_0$.
 
\begin{itemize} 
 
\item {\it Characterization of scaling properties.} 

Given a distribution of points, 
the first main question concerns the possibility
of defining a physically meaningful average density.
In fractal-like systems such a quantity depends
on the size of the sample, and it does not
represent  a reference value, as in the case
of an homogeneous distribution. 
Basically a system cannot be homogeneous
below the scale of the maximum void
present in a given sample. 
However the complete statistical characterization
of highly irregular structures is the objective of
Fractal Geometry\cite{man77}.

The major problem from the point of view 
of data analysis is to use statistical methods 
which are able to properly characterize scale 
invariant distributions, and hence which are 
also suitable to characterize an eventual 
crossover to homogeneity.  
Our main contribution \cite{pie87,cp92,slmp98}, 
in this respect, has been to clarify that the usual  
statistical methods (correlation function, 
power spectrum, etc.) are based on the assumption of  
homogeneity and hence are not appropriate 
to test it. Instead, we have introduced and developed 
various statistical tools which are able 
to test whether a distribution is  
homogeneous or fractal, and to correctly characterize 
the scale-invariant properties. 
Such a discussion is clearly relevant also 
for the interpretation of the properties 
of artificial simulations. The agreement about  
the methods to be used for the analysis of future 
surveys such as the Sloan Digital Sky Survey (SDSS) 
and the two degrees Fields (2dF) is clearly a fundamental  
issue. 

Then, if and only if the average density is found to be
not sample-size dependent, one may study the 
statistical properties of the fluctuations 
with respect to the average density itself.
In this second case one can study 
basically two different length scales.
The first one is the homogeneity scale ($\lambda_0$),
which  defines
the scale beyond which the density fluctuations
become to have a small amplitude with respect to the
average density ($\delta \rho < \rho$).
 The second scale is related to the 
typical length scale of the structures of 
the density fluctuations, and, according to the terminology
used in statistical mechanics \cite{perezmercader}, it is called 
correlation length $r_c$. Such a scale has nothing
to do with the so-called "correlation length" 
used in cosmology and corresponding to
the scale $\xi(r_0)=1$\cite{pee80}, which is instead
related to $\lambda_0$ if such a scale
exists.
 
\item {\it Implication of the fractal structure up 
to scale $\lambda_0$}. 
 
 The fact that galactic structures 
are fractal, no matter what is the homogeneity scale $\lambda_0$, 
 has deep implication on the interpretation 
of several phenomena such  
as the luminosity bias, the mismatch 
galaxy-cluster, the determination of the average  
density, the separation of linear and  
non-linear scales, etc.,  
and on the theoretical concepts used 
to study such properties. 
We discuss in  detail some of these points.
 We then review some of the main consequences 
of the power law behavior of the galaxy number
density, by relating various observational
quantities (e.g. $r_0$, $\sigma_8$, $\Omega$, etc.) 
to the length scale $\lambda_0$. 

We also note that the properties 
of dark matter are inferred from the ones of visible 
matter, and hence they are closely related. 
If now one observes different  
statistical properties for galaxies and clusters, 
this necessarily implies a change of perspective 
on the properties of dark matter.

\item {\it Determination 
of the homogeneity scale $\lambda_0$. } 
 
This is, clearly, a very important point  
which is at the basis of the understanding 
of galaxy structures and more generally 
of the cosmological problem. We distinguish 
here two different approaches: direct tests and  
indirect tests. By direct tests, we mean the determination 
of the conditional average density in three dimensional surveys, 
while with indirect tests we refer to other possible analyses, 
such as the interpretation of angular surveys, the  
number counts as a function of magnitude or of distance  or, 
in general, the  
study of non-average quantities, i.e. when the fractal
dimension is estimated without making an average 
over different observes (or volumes).
While in the first case one is able to have a  
clear and unambiguous answer from the data, in the second 
one is only able to make some weaker  
claims about the compatibility of the data 
with a fractal or a homogeneous distribution. 
For example   the papers of Wu et al\cite{rees99}
and  Nusser \& Lahav\cite{nl00} 
mainly concern with compatibility arguments,
rather than with direct tests.
However, also in this second case, it is possible to understand  
some important properties of the data, and to 
clarify the role and the limits of some underlying  
assumptions which are often used without a  
critical perspective. 
We do not enter here in the details of the discussion
about real data (see e.g. Joyce et al. 1999, Wu et al. 1999),
 however
in the last section we consider separately the case (i) 
$\lambda_0 \approx 50 \hmp$,
 (ii) 
$\lambda_0 \approx 300 \hmp$ and 
  (iii) 
$\lambda_0 \approx 1000 \hmp$, 
briefly discussing the main theoretical consequences.
\end{itemize} 
 
\section{Characterization of scaling properties}

In this section we describe in detail the correlation
properties  of a fractal
distribution of points having eventually   a
crossover towards homogeneity (see Gabrielli \& Sylos Labini, 2000
for a more exhaustive discussion of the subject),
 as the distribution of galaxies is thought
to be (see Sylos Labini, Montuori \& Pietronero, 1998  and 
Wu, Lahav \& Rees,  1999  for two opposite views on the matter).

Let 
\be
\label{corr1b}
n(\vec{r}) = \sum_i \delta(\vec{r} - \vec{r}_i)
\ee
be the number density of points in the system (the index $i$ runs 
over all the points) and let us suppose to have an infinite system.
If the presence of an object at the point $\vec{r}_1$ 
influences the probability of finding another object 
at $\vec{r}_2$, 
these two points are correlated. Hence  there is a correlation
at the scale distance $r$ if 
\be
\label{corr1}
G(r) = \langle n(\vec{0})n(\vec{r})\rangle   \ne \langle n\rangle  ^2  
\ee
where we average over 
all occupied points of the system  chosen as origin and on the
total solid angle,  supposing statistical isotropy.
On the other hand, there is no correlation if
\be
\label{corr2}
G(r) = \langle n\rangle  ^2.
\ee

\subsection{Homogeneity scale and correlation  length}

The proper definition of  $\lambda_0$, the {\it homogeneity scale},
is  the length scale beyond which the average density
becomes to be well-defined, i.e. 
there is a crossover towards homogeneity with a flattening of $G(r)$.
The length-scale $\lambda_0$ is related to  the typical dimension
of the largest voids in the system.
On the other hand, the {\it correlation length} $r_c$  
separates correlated regimes 
of the fluctuations with respect to the average density
from
uncorrelated ones,
 and it can be defined only if a crossover towards homogeneity is 
shown by the system, i.e. if $\lambda_0$ exists\cite{perezmercader}.
In other words $r_c$ defines the   organization
in geometrical structures of the fluctuations 
with respect to the average density. Clearly
$r_c > \lambda_0$:
only if the average density can be defined 
one may study the correlation length of the fluctuations
from it.
In the case  $\lambda_0$ is finite 
and then $\langle n \rangle >0$, in order 
to study the correlation properties of the fluctuations around the
average and then the behaviour of $r_c$, we can introduce the 
correlation function
\be
\label{corr5} 
\xi(r)  =  \frac{ \langle n(0)n(r)\rangle   - \langle n\rangle^2} 
{\langle n\rangle^2} \; .
\ee
In the case  of a fractal distribution, the average
density $\left<n\right>$ in the {\bf infinite system} is zero, then 
$G(r)=0$ and $\lambda_0 = \infty$
and consequently $\xi(r)$ is not defined.
In this case the only well defined quantity characterizing the
two point correlations is the function 
$\Gamma(r)$ \cite{pie87,cp92}:
\be
\label{corr4} 
\Gamma(r) = \lim_{R_s\rightarrow \infty} 
\frac{\langle n(r)n(0) \rangle_{R_s}}{\langle n \rangle_{R_s}}
\ee
where $R_s$ is the size of the a generic finite sample of the system, 
$\left<...\right>_{R_s}$ indicates the 
average over all the points of the sample as origins,
hence $\langle n \rangle_{R_s}$ is the average density of the sample.
This function measures the average density of points 
at a distance $r$ from another occupied point,
and this is the reason why it is called
the conditional average density \cite{pie87}.
Obviously in the case of a distribution
for which $\lambda_0$ is finite $\Gamma(r)$ provides the
same information of $G(r)$, i.e. it 
characterizes the correlation properties for $r < \lambda_0$
and the crossover to homogeneity.

A very important point
is represented by the kind of information 
about the correlation properties
of the {\bf infinite system}
which  can be extracted from the analysis
of a {\bf finite sample} of it.
In Pietronero (1987) 
it is demonstrated that even 
in the super-correlated 
case of a fractal the estimate of $\Gamma(r)$ 
extracted from a finite sample, is 
not dependent on the sample size $R_s$, 
providing a good approximation of 
that of the whole system. Clearly this is true a part from statistical 
fluctuations due to the fact that in a finite sample the average over the all
possible origins is an average over a finite number of points, while
 in the
global infinite system the average is over an infinite number of points.
 In fact,
 $\Gamma(r)$ extracted from a sample can be written in the following way:
\be
\label{and0}
\Gamma(r) = \frac{1}{N} \sum_{i=1}^{N} \frac{1}{4 \pi r^2 \Delta r}
\int_{r}^{r+\Delta r} n(\vec{r}_i+\vec{r}')d^{3}r',
\ee
where $N$ is the number of points in the sample,
$n(\vec{r}_i+\vec{r}')$ is the number of
points in the volume element $d^3r'$ around the point 
$\vec{r}_i+\vec{r}'$ and $\Delta r$
is the thickness of the shell at distance $r$ from the point at $\vec{r}_i$.

Therefore, from an operative point of view,
having a finite sample of points (e.g. galaxy catalogs),
the first analysis to be done concerns the determination of 
$\Gamma(r)$ of the sample itself. Such a measurement
 is necessary to distinguish
between the two cases:
(1) a crossover towards
 homogeneity in the sample shown by
 a flattening of $\Gamma(r)$,
and hence an estimate of $\lambda_0<R_s$ and $\langle n \rangle$;
(2) a continuation of the fractal behavior.
Obviously only in the case (1), it is 
physically meaningful to study  the correlation function $\xi(r)$ 
(Eq.\ref{corr5}), and extract from it the length scale
$r_0$ ($\xi(r_0) = 1$), which is related
to the intrinsic homogeneity scale $\lambda_0$.
The functional behavior of $\xi(r)$ with distance 
gives instead information on the correlation length of
the density fluctuations.


\subsection{The case of a fractal distribution ($R_s\ll\lambda_0$)}

Hereafter we   study the three-dimensional case,
 i.e. $d=3$, and we suppose that 
the sample is a sphere of radius $R_s$. Obviously, 
this choice is not a restriction.

Let us analyze the case  
$R_s\ll\lambda_0<r_c$. This is the so called ``fractal'' case,
and it is compatible with both the situation of $\lambda_0$ finite,
but $R_s\ll\lambda_0$ (a sample-size which is smaller than 
the homogeneity scale), or the
situation in which $\lambda_0\rightarrow \infty$, i.e. the 
case of a fractal distribution
at any scale.
 
It is simple to show\cite{pie87,cp92,slmp98} 
that in this case (and in a spherical sample),
Eq.\ref{and0} becomes
\be
\label{fra4}
\Gamma(r) =\frac{BD}{4 \pi} r^{D-3}
\ee
with $B= N/ R_s^D$. Note that $B$ is independent on the sample size:
 in fact, by changing 
$R_s$, $N$ in average scales as $R_s^D$. This shows the 
aforementioned assertion
that $\Gamma(r)$ is practically independent on the sample-size. 
On the other hand, it is possible
to show  that $B$ 
is related approximatively to 
the average distance between nearest neighbors points 
in the system \cite{slm98}:
\be
\label{fra6}
\ell   \approx \left(\frac{1}{B}\right)^{\frac{1}{D}} \Gamma_e
\left(1 + \frac{1}{D} \right)                                    
\ee
where $\Gamma_e$ is the Euler's gamma-function.


\subsection{The ``standard'' correlation function 
for a fractal distribution}

As already mentioned, in the fractal case ($R_s\ll\lambda_0$), 
the sample estimate of the 
homogeneity scale, through the value of $r$ for which the 
sample-dependent correlation function 
$\xi(r)$ (given by Eq.\ref{and0b}) is equal to $1$, is meaningless:
This estimate is the  so-called ``correlation length''
$r_0$ \cite{pee80} in the standard approach of 
cosmology. As we discuss below,
 $r_0$ has nothing to share with the {\it true} correlation length $r_c$.
 Let us see why $r_0$ is unphysical in the case $R_s\ll\lambda_0$.
$\xi(r)$ is given operatively by
\be
\label{and0b}  
\xi(r)=\frac{\langle n(r) n(0) \rangle_{R_s}}{\left<n\right>_{R_s}^2} -1 
=
\frac{\Gamma(r)}{\left<n\right>_{R_s}} -1 \; .
\ee
The basic point in the present discussion\cite{pie87},
is that the mean density of the sample, $\langle n \rangle_{R_s} $,
used in the normalization of  $\xi(r)$, is not an intrinsic quantity 
of the system,
but it is a function of the finite size $R_s$ of the sample.

In fact, from Eq.\ref{fra4},   
the expression of the $\:\xi(r)$ of the sample in the case of
fractal distributions is \cite{pie87}
\be
\label{xi3}
\xi(r) = 
\frac{D}{3} \left( \frac{r}{R_s} \right)^{D-3} -1 \; .
\ee
From Eq.\ref{xi3} it follows that $\:r_0$ 
is a linear function of the sample size $\:R_{s}$
\be
\label{xi4}
r_{0} =\left(\frac{D}{6}\right)^{\frac{1}{3-D}}R_{s}
\ee
and hence it is a spurious quantity without  physical meaning but it is
simply related to the sample's finite size.

We note that the amplitude of $\Gamma(r)$ (Eq.\ref{fra4})
is related to the lower
cut-off of the fractal $\ell$ by Eq.\ref{fra6}, while the amplitude of $\xi(r)$
is related to the upper cut-off (sample size $R_s$) of the distribution. 
This crucial difference has never been appreciated  appropriately.

Finally we stress that in the standard analysis of galaxy catalogs
the fractal dimension is estimated by fitting $\xi(r)$ with a power law, 
which instead, as one can see from
 Eq.\ref{xi3}, it is power law only for  $r \ll r_0$ (or $\xi \gg 1$).
For larger distances there is a clear deviation
from the power law behavior due to the definition of $\xi(r)$.
Again this deviation is due to the finite size of the observational 
sample and does not correspond to any real change
in the correlation properties. It is easy to see that, if 
one estimates the exponent 
at distances $r \ltapprox r_0$, one 
systematically obtains a higher value 
of the correlation exponent due to the break 
of $\xi(r)$ in a log-log plot. 
To illustrate more clearly this we compute
the log derivative of Eq.\ref{xi3} with respect to $\log(r)$, indicating $D-3$ 
with $\gamma$ and its estimate with $\gamma'$:
\be
\label{xi6}
\gamma'=\frac{d(\log(\xi(r))}{d\log(r)} =
\frac{2 r_0^{\gamma} r^{-\gamma}} 
{2 r_0^{\gamma} r^{-\gamma} -1} \, \gamma \; ,
\ee
where $r_0$ is defined by Eq.\ref{xi4}.
 The tangent
to $\xi(r)$ at $r=r_0$ has a slope $\gamma'= 2\gamma$.
This explain why it has been found in galaxy and cluster 
catalogs that $\gamma \sim 2$  by the $\xi(r)$ analysis
\cite{dp83,dav97,rees99}
instead of $\gamma \sim 1$ found with the $\Gamma(r)$ analysis 
\cite{slmp98}.


\subsection{The case of a fractal
 distribution with  a crossover to homogeneity ($R_s\gg\lambda_0$)}

Let us now analyze the   case of a fractal
with a crossover to homogeneity.
In Coleman \& Pietronero (1992)  
a very simple approximation has been 
used to describe such a situation which we discuss 
in more detail below.

By defining
$\left<n\right>_{R_s}=N/V$ with $V=4\pi R_s^3/3$, 
it is simple to see\cite{cp92}  that
the behavior of $\Gamma(r)$ 
in our sample is fractal (i.e. $\Gamma(r)$ is a power law) 
up to a certain distance $\lambda_0$,
and then it flattens, becoming homogeneous at scales
 $R_s > r \gg \lambda_0$:
\be 
\left\{ 
\begin{array}{l} 
\label{frao1}
\Gamma(r)= \frac{DB}{4\pi} r^{D-3} \; \; \mbox{for}
\; \;\ell \le r \ll \lambda_0
\\
\\
  \Gamma(r) 
\simeq \left<n\right>_{R_s}
  \; \; \mbox{for}
  \; \;\lambda_0 \ll r \le R_s \,,
  \end{array} \right.
 \ee 
where $\left<n\right>_{R_s}$ is the estimation of the 
average density in the sample of size $R_s$. That is,
$\left<n\right>_{R_s}$ does
not depend on $r$ if $\lambda_0 \ll r \ltapprox R_s$,
apart small amplitude fluctuations.
In Eq.\ref{frao1} the detailed approach to homogeneity
depends on the specific properties of the fluctuations around the average
density, i.e. it is determined by $r_c$.
Hence, the statistical properties of the density fluctuations
determine how good is the estimation of the 
average density throught  $\left<n\right>_{R_s}$.
 
From the definition of the function $\xi(r)$  we can find\cite{gsl00}
\be
\label{frao1b}
\xi(r) \approx 
\left(\frac{r}{\lambda_0}\right)^{D-3}f\left(\frac{r}{r_c}\right) \; .
\ee
Note that the amplitude of $\xi(r)$ is determined
 by the homogeneity scale $\lambda_0$
which has been previously extracted from $\Gamma(r)$, and that in this
approximation
$r_0 \sim \lambda_0 \;$ .
The function $\xi(r)$ characterizes the correlations among the 
fluctuations of the distribution
with respect to the average density.
It is important to clarify that these fluctuations must be both 
positive and negative, in fact
the integral over the whole sample of Eq.\ref{and0b} must be $0$.
Hence the function $f\left(\frac{r}{r_c}\right)$ must be oscillating, and 
in the case  $\lambda_0 < r_c , R_s$, it should present
an exponential cut-off at $r \approx r_c$. 
We have that, when 
$\xi(\lambda_0) \simeq 1$, the density fluctuations begin to become 
small with respect to the 
average density $\langle n \rangle$, but if $\lambda_0<r<r_c$
they are still well correlated among them. Only for $r\gg r_c$ 
the fluctuations are not
correlated.

Let us now consider the case  $\lambda_0\ll R_s<r_c$. 
This situation is compatible with
the following two situations: $r_c$  finite, but larger than $R_s$,
and the case $r_c\rightarrow\infty$.
In both cases, $\xi(r)$ of our sample should be a power law
modulated by an oscillating function $g(r)$ which describes
the positive and negative fluctuations with respect
to the average density,
\be
\label{and4bis}
\xi(r) = \left(\frac{r}{\lambda_0}\right)^{-\gamma} g(r) \; .
\ee
In such a situation the (positive and negative) fluctuations
from the average density are of all sizes and they do not
have any intrinsic characteristic scale: this is a critical system 
(see Gaite et al., 1999  for a more detailed discussion). The only 
intrinsic scale of the system is then $\lambda_0$,
the length-scale beyond which $\Gamma(r)$ flattens and the fluctuations
are small with respect to the average.

Let us suppose to be in the case  in which  $\Gamma(r)$ 
flattens at a certain
$\lambda_0\ll R_s$. We can then evaluate the correlation
function $\xi(r)$ of the sample via Eq.\ref{and0b}. 
At this point we can clarify 
how to interpret the eventual cut-off shown by $\xi(r)$.
\begin{itemize}
\item If the cut-off scale 
is well below $R_s$, we can be sure that it is a good
estimate of the intrinsic correlation length $r_c$;
\item if the cut-off is at a scale $r\simeq R_s$ we can
 have two cases: it represents 
an ``intrinsic'' cut-off with 
$r_c\simeq R_s$, or it is only a finite-size effect due to
the fact that from Eq.\ref{and0b} $\left|\xi(R_s)\right|=0$.
In order to distinguish between these two possibilities, it
 is necessary to increase 
the sample size and to look at the behavior of the cut-off scale. 
If it increases proportionally to $R_s$, then  it is a finite size effect.
Otherwise  if it
does not change, it represents the estimate of the intrinsic
 correlation length $r_c$.
\end{itemize}

In Fig.\ref{gamma} we show two possible
behaviours of the flattening of $\Gamma(r)$,
while in Fig.\ref{xi} it is shown the corresponding
$\xi(r)$ (we neglect for simplicity the oscillating term
which must be present, and we have considered  the situation
$R_s\rightarrow\infty$).
 
\bef 
\epsfxsize 10cm
\centerline{\epsfbox{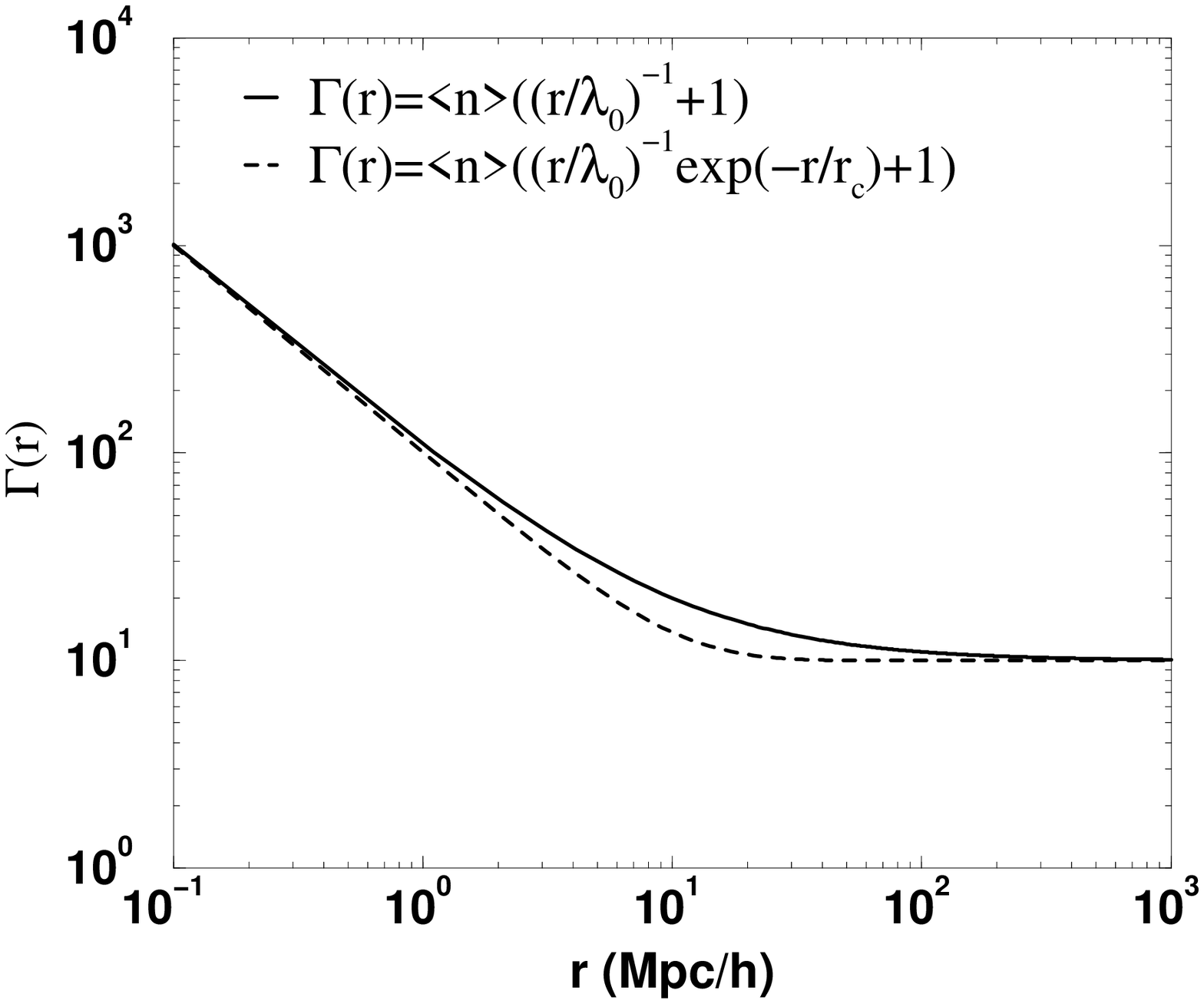}}
\caption{\label{gamma} The conditional average density $\Gamma(r)$
for a distribution which
has a power law behavior at small scales ($r < \lambda_0$) , followed 
by a transition to homogeneity at the  scale $\lambda_0 = 10 \hmp$. 
The behavior of the flattening
depends on the correlation properties
of the density fluctuations, i.e. on the functional behavior of
$\xi(r)$. The dotted line corresponds to a system which has a finite
correlation length $r_c = 30 \hmp$, while the solid line 
describes a system whose density fluctuations present
correlation over all scales. From the $\Gamma(r)$-analysis alone it is
possible to compute $\lambda_0$ but not $r_c$.
}
\eef
\bef 
\epsfxsize 10cm
\centerline{\epsfbox{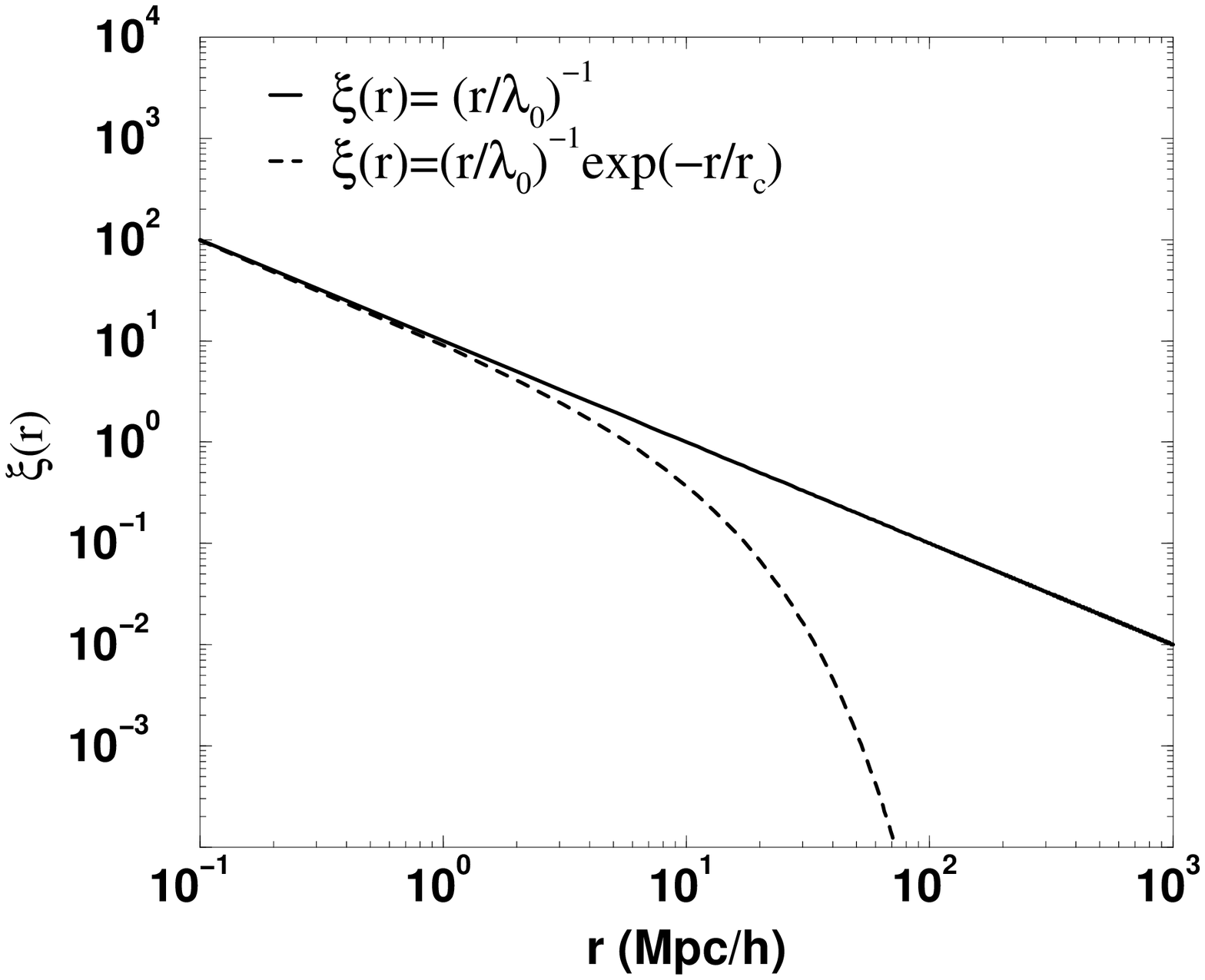}}
\caption{\label{xi} The $\xi(r)$ correlation function
for the distributions shown in the previous figure.
With this analysis it is possible to 
compute the correlation length $r_c$ (finite or infinite) of the density
fluctuations. The dotted line corresponds to an exponential decay,
and hence to a finite value of the correlation length ($r_c = 30 \hmp$)
while the solid line corresponds to an infinite correlation
length, and hence to a power law behavior. The length scale
at which $\xi(r_0)=1$ gives a reliable estimation of $\lambda_0$.
}
\eef

\subsection{About the amplitude of $\xi(r)$}

We note that if $\lambda_0 \ll R_s$, $\lambda_0$
has nothing to share with questions like  ``which is the typical size 
of structures in the system?'' or 
``up to which length-scale the system is clusterised?''.
The answer to this question is strictly related to $r_c$ and not to 
$\lambda_0$. 
The length scale
 $r_c$   characterizes the  distance over which two different points 
are correlated (clusterised)\cite{perezmercader}. 
In fact, this property is not related to how large are the 
fluctuations with respect to the average ($\lambda_0$),
 but to the length extension of their 
correlations ($r_c$). 

To be more specific, 
let us consider a fixed set of density fluctuations.
They can be superimposed to
different values of a uniform density background.
The larger is this background the lower $\lambda_0$, but obviously
the   length scale 
of the correlations ($r_c$) among these fluctuations is not changed, i.e.
they are clusterised independently of the background.

One can see\cite{gsl00,gsld00} that a linear amplification of $\xi(r)$
such that
\be
\label{and7}
\xi'(r)= A \xi(r)
\ee
doesn't change $r_c$ (which can be finite or infinite)
but only $\lambda_0$, i.e. if $A>1$
we need  larger subsamples to have a good 
estimation of $\langle n \rangle$,
but the characteristic length 
(correlation length) of the structures is not changed.

\subsection{Homogeneity scale and the size of voids}

Basically $\lambda_0$ is related to the maximum 
size of voids: the average density will be constant, 
at least, on scales larger than the maximum 
void in a given sample. Several authors have  
approached this problem by looking at  
voids distribution. For example 
El-Ad and Piran (1997) have shown that the SSRS2  
and IRAS 1.2 Jy. redshift surveys are dominated  
by voids: they cover the $\sim 50 \%$ of the volume. 
Moreover the two samples show very similar  
properties even if the IRAS voids 
are $\sim 33 \%$ larger than SSRS2 ones because they  
are not bounded by narrow angular limits as the SSRS2 voids.  
The voids have a scale of at least $\sim 40 \div 50 \hmp$ 
and the largest void in the SSRS2 sample has a 
diameter of $\sim 60 \hmp$, i.e. comparable 
to the Bootes void. The problem is to understand whether 
such a scale has been fixed by the samples' volume, 
or whether there is a tendency  not to find larger 
voids: in this case one would have a (weaker evidence) for the  
homogeneity scale. In any case, we note that 
the homogeneity scale cannot be smaller 
than the scale of the largest void found in these samples 
and that one has to be very careful when comparing the size of 
the voids to the effective depth of catalogs. For example  
in the Las Campanas Redshift Survey, even if  
it is possible to extract sub-samples limited at $\sim 500 \hmp$, 
 the volume of space investigated is not so 
large, as the survey is made by thin slices. In  
such a situation a definitive answer to the dimension 
of the of voids, and hence to the existence of the homogeneity 
scale, is rather difficult and uncertain. 


\subsection{Luminosity Bias} 
 
We would like to stress again that, even if the  
fractal behavior breaks at a certain scale $\lambda_0$, 
the use of $\xi(r)$ is in anyhow  inconsistent at scales 
smaller than $\lambda_0$.  We illustrate below
an example of the confusion due to the use 
of $\xi(r)$ when $r]ll \lambda_0$.

From the use of the $\xi(r)$ analysis, it has been 
found that $r_0$ is different in different volume  (hereafter VL) 
samples.
In particular it has been found \cite{ben96} that 
deeper is the VL sample, larger is the value 
of $r_0$. As the deeper VL samples
contain brighter galaxies, this fact has been
interpreted as a real physical phenomenon,
leading to the idea that more brighter galaxies
are more strongly clustered than fainter ones,
in view of their larger correlation amplitude:
this is the so-called  luminosity segregation phenomenon
\cite{dav88,par94}.
In other words, the fact that the giant galaxies are "more clustered"
than
the dwarf ones,
i.e. that they are located in the peaks of the density field,
has given rise to the proposition
that larger objects may correlate up to larger
length scales and that the amplitude of 
$\:\xi(r)$ is larger
for giants than for dwarfs one. The deeper VL samples
contain galaxies which are in average
brighter than those in the VL samples with
smaller depths. As the brighter galaxies should have
a larger correlation length the shift of $r_0$ in different samples
can be related, at least partially,
with  the phenomenon of luminosity segregation.

As previously discussed there are two 
problems with such a model: 
(i) The amplitude of $\xi(r)$ in an 
homogeneous distribution, does not give any
information about the clustering "strength".
It is instead related to the local
amplitude of the fluctuations with respect to the average density.
(ii) The amplitude of $\xi(r)$ has a physical meaning only
in the case $\lambda_0$ is found to be finite
and smaller than the sample's size. This 
is clearly not the case up, at least, to $\sim 50 \hmp$.

A natural explanation of the scaling of $r_0$ is
then the fractal behavior of galaxy distribution,
and more specifically the fact that 
$r_0$ is a fraction of the sample's size 
in the fractal case. The fact that
giant elliptical galaxies are located 
in the core of rich clusters, and other
morphological properties of this kind,
can be naturally related to the 
multifractal properties of matter distribution\cite{cp92,slp96}.
In such a case, bright galaxies are more strongly
clustered than fainter ones in view of the fact
that their fractal dimension is smaller\cite{gsl00}.

\subsection{Power Spectrum of density fluctuations} 
 
The problems  
with the standard correlation analysis also show that the  
properties of fractal correlations have not been really appreciated. These  
problems are actually far more serious and fundamental than mentioned, 
for example, by  
Landy \cite{landy99}  
and the idea that they can be solved by simply taking the Fourier  
transform is once more a proof of the superficiality of the discussion.  
We have extensively shown \cite{sla96,slmp98} that 
the power spectrum of the density fluctuations 
has the same kind of problems which $\xi(r)$ has, 
because it is normalized to the average density as well. 
The density contrast $\delta(r)=\delta \rho(r)/ \langle \rho \rangle$ 
is not a physical quantity unless the average density 
is demonstrated to exist.  
More specifically, like in the case of $\xi(r)$, the power spectrum 
(Fourier Transform of the correlation function) 
is affected by finite size effects at large scale:  
even for a fractal distribution the power spectrum 
has not a power law behavior but it shows 
a large scale (small $k$) cut-off which  
is due to the finiteness of the sample \cite{sla96}. 
Hence the eventual detection of the turnover of the power 
spectrum, which is expected in CDM-like models  
to match the galaxy clustering to the anisotropies 
of the CMBR, must be  considered a finite size effect, 
unless a clear determination of the average density 
in the {\it same sample} has been done.

Essentially all the currently
elaborated models of galaxy formation 
\cite{pee93} 
{\it assume large scale homogeneity} and 
predict that the galaxy
power spectrum (hereafter PS), 
which  is {\it the PS  of the density contrast},
decreases both toward small scales and toward large
scales, with a turnaround somewhere in the middle, at a scale $\lambda_f$
that can be taken as separating ``small'' from ``large'' scales. 
Because of the homogeneity assumption, the PS    amplitude
should be independent on the survey scale, any  residual 
variation being attributed to luminosity bias (or to the
fact that the survey scale has not yet reached the homogeneity scale).
However, the crucial clue to this
picture, the firm determination of the 
scale $\lambda_f$, is still missing, although
some surveys do indeed produce a  turnaround
 scale around 100 $\hmp$
\cite{be94,fe94}. 
Recently, the CfA2  survey 
analyzed by \cite{par94} 
(hereafter PVGH) 
(and confirmed by SSRS2 \cite{dac94} 
- hereafter  DVGHP), showed a $n=-2$ slope up to $\sim 30 \hmp$,
a milder $n\approx -1$ slope  up to 200 $\hmp$, and some tentative
indication of flattening on even larger scales. PVGH also find
that deeper subsamples have higher power amplitude,
i.e. that the amplitude scales with the sample depth.

{\it In the following  we argue  that both features, bending and scaling,
are a manifestation of 
the finiteness of the survey volume, and that they
cannot be
interpreted as the convergence to homogeneity, nor to a PS   
flattening.}
The systematic effect of the survey finite size is in fact
to suppress
power at large scale, mimicking a real flattening.
Clearly, this effect occurs whenever galaxies have  not 
a    correlation scale much larger than the survey size, and it has
often been studied in 
the context of standard scenarios \cite{it92,col94}.
 We push this argument further, by showing that
even a fractal distribution of matter, 
which never reaches  homogeneity, shows a sharp flattening
and then a turnaround. Such 
features are  partially corrected, but not quite eliminated,
  when the correction proposed by \cite{pn91}  is applied to the data.
 We show also
 how the amplitude of the
PS    depends on the survey size as long as 
the system shows long-range correlations.

The standard  PS    (SPS) 
measures directly the contributions of different scales to the galaxy
density contrast $\:\delta\rho/\rho$.
It is clear that the density contrast, 
and all the quantities based on it,  is meaningful only when one can define
a constant density, i.e. reliably identify
the sample density with   
the average density of all the Universe.
In other words in  {\it the SPS analysis 
one assumes that the survey volume is large enough
to contain a  homogeneous sample.} 
When this is not true, and we argue that is 
indeed an incorrect assumption
in all the cases investigated so far, a false interpretation of the results may
occur, since both 
the shape and the amplitude of the PS    (or correlation
function) depend on the
survey size.

Let us recall the basic  notation of the PS    analysis.
Following Peebles \cite{pee80}  we imagine that the Universe is periodic
in a volume $\:V_{u}$, with $\:V_{u}$ much 
larger than the (presumed) maximum
homogeneity scale. The survey volume $V\in V_u$
contains $\:N$ galaxies at positions $\:\vec{r_i}$,
and the galaxy density contrast is
\be
\label{eps4}
\delta(\vec{r}) = \frac{n(\vec{r})}{\hat n} -1 
\ee
where it is assumed that exists a 
well defined constant density $\hat n$, obtained
averaging over a sufficiently large scale.
The density function
$\:n(\vec{r})$ 
can be described by a sum of delta functions:
$~n(\vec{r}) = \sum_{i=1}^{N} \delta^{(3)} 
(\vec{r}-\vec{r_{i}})\,.~$
Expanding the density contrast in its Fourier components we have 
\be
\label{eps7}
\delta_{\vec{k}} = \frac{1}{N} \sum_{j \epsilon V} 
e^{i\vec{k}\vec{r_{j}}} - W(\vec{k})\,,
\ee
where
\be
\label{eps7b}
~W(\vec{k}) = \frac{1}{V} \int_V d{\vec{r}} W(\vec{r})
 e^{i\vec{k}\vec{r}}\,~
\ee
is the Fourier transform of the survey window $W(\vec{r})$, 
defined to be unity inside the survey region, and zero outside.
If $\xi(\vec{r})$ is the correlation function of the galaxies
($\xi(\vec{r}) = <n(\vec{r})n(0)>/\hat n^2 -1$),
the true PS $\:P(\vec{k})$ is defined as
the Fourier conjugate of the 
correlation function $\xi(r)$.
Because of isotropy the PS can be simplified to
\be
\label{eps12}
P(k)  =4\pi \int \xi(r)  \frac{\sin(kr)}{kr} r^{2}dr\,.
\ee
The  variance of $\:\delta_{\vec{k}}$ 
is \cite{pee80,pn91,fis93} \be
\label{eps9}
<|\delta_{\vec{k}}|^{2}> = \frac{1}{N} + \frac{1}{V}\tilde P(\vec{k})\,.
\ee
The first term is the usual additional shot noise term
while the second is the true PS convoluted with a 
window function 
which describe the geometry of the sample (PVGH)
\be
\label{eps9b}
\tilde P(\vec{k})= 
{V\over (2\pi)^3} \int
<|\delta_{\vec{k'}}|^{2}>
|W(\vec{k}-\vec{k'})|^2
d^3 \vec{k'}
\,.
\ee
\be
\label{eps9bb}
\tilde P(\vec{k}) = \int d\vec{k'} P(\vec{k'}) F(\vec{k}-\vec{k'}) \,,
\ee
with 
\be
\label{eps9c}
F(\vec{k}-\vec{k'}) = \frac{V}{(2\pi)^3} |W(\vec{k}-\vec{k'})|^2\,.
\ee
We apply now this standard analysis to a fractal distribution.
We recall the  expression 
of   $\:\xi(r)$ in this case is 
\be
\label{eps16}
\xi(r) = [(3-\gamma)/3](r/R_{s})^{-\gamma} -1\,,
\ee
 where  $\:\gamma=3-D$. 
 A key point of our discussion is
that  that
on scales larger that $R_s$ the $\xi(r)$ cannot be calculated without
making assumptions on the 
distribution outside the sampling volume.

As we have already mentioned, in a fractal
quantities like 
$\xi(r)$ 
are  scale dependent: in particular both
the amplitude and the shape of $\xi(r)$ depend 
the survey size.
It is clear  that the same kind of 
finite size effects  are also present when computing the SPS, so that 
it is very dangerous to identify real physical features induced
from the SPS analysis without first a firm determination of the
homogeneity scale.
 
The SPS for a fractal distribution 
 model described by
 Eq.\ref{eps16}
inside a sphere of radius $R_s$ is
\be 
\label{eps19}
P(k) =  \int^{R_{s}}_{0} 
   4\pi \frac{\sin(kr)}{kr} \left[ \frac{3-\gamma}{3} 
\left(\frac{r}{R_{s}}\right)^{-\gamma} -1\right] r^{2}dr=
\frac{a_k(R_s) R_{s}^{3-D}}{k^{D}}-\frac{b_k(R_s)}{k^{3}}\,.
 \ee
Notice that the integral has to be evaluated inside $R_s$
because we want to compare $P(k)$  with its {\it estimate}
 in a finite size spherical survey of scale $R_s$. 
 In the  general case, we must deconvolve the
 window contribution 
 from $P(k)$; $R_s$ is then a characteristic window scale. 
Eq.\ref{eps19} shows the two scale-dependent features of the PS. First,
the amplitude of the PS 
depends on the sample depth.
Secondly,
the shape of the PS 
is characterized by two  scaling regimes:
the first one, at high wavenumbers, 
is related to the fractal dimension of the 
distribution in real space, 
while the second one arises only because of 
the finiteness of the sample.
In the case of $\:D=2$  in Eq.\ref{eps19} one has:
\be
\label{eps22}
~a_k(R_s) = \frac{4\pi}{3} (2+\cos(kR_{s}))\,
\ee
and
\be
\label{eps23}
~b_k(R_s) = 4\pi \sin (kR_{s})\,.
\ee
The PS is then a power-law with exponent 
$\:-2$ at high wavenumbers, 
it flattens at low wavenumbers and reaches a maximum at
$k\approx 4.3/R_s$, i.e. at a scale $\lambda \approx 1.45 R_s$.
The scale at which the transition occurs 
is thus related to the sample depth. 
In a real survey, things are complicated by the window function,
so that the flattening (and the turnaround) scale can only be determined
numerically.

In practice one has several complications. First, the survey
in general is not spherical. This introduces a coupling with
 the survey window
which is not easy to model analytically. 
For instance, we found that windows of small
angular opening shift to smaller scales
the PS turnaround. 
This is analogous to what happens with the correlation function
of a fractal: when it is calculated in small angle surveys, the
correlation length $r_0$ decreases. Second, the observations
are in redshift space, rather than in real space. 
The peculiar velocities generally make steeper the PS    slope 
\cite{fis93}
 with respect to the real space. Third, in a fractal
the intrinsically high level of fluctuations makes hard
a precise comparison with the theory when the fractal under study
is composed of a relatively small number of points.

\section{Implications for cosmology}

We now consider some   implications
for cosmology of the scaling properties of
galaxy distribution, up to a lenght scale $\lambda_0$.
For example,
we consider more specifically   $\lambda_0 \approx 50 \hmp$.

\subsection{Estimation of the average luminosity and mass density}

From the studies of Large Scale Structures (LSS)
of galaxies and galaxy clusters, one would like to estimate
the average density of visible matter and then
to infer the one of the whole (visible plus dark,
i.e. all the matter in clusterised objects) 
matter distribution.
While for the first we have direct estimations,
for the second we have only indirect methods, especially at large scales,
based on some assumptions, which can be tested by looking at the
distribution of what is observable, i.e.  visible matter.

We briefly describe how to do such a measurement
in galaxy redshift catalogs directly, from the knowledge
of galaxy positions and luminosity (for a more detailed
discussion see Sylos Labini, 2000). In this
case, and by measuring the   Mass-to-Luminosity
ratio, one  can infer, from the average  luminosity
density, the average  mass density.
The new point we address more specifically is that
galaxies are fractally distributed up to a certain
crossover scale $\lambda_0$. As there is still some
controversy about the value 
of $\lambda_0$ \cite{slmp98,rees99,chown99,jmsl99,martinez99} we 
give the estimation as a function
of $\lambda_0$.
We stress that the way this estimation is performed,
is substantially different from the usual one \cite{pee93},
because in such a case the fractal behavior is not considered
at all, and one assumes a perfect homogeneous 
distribution at relatively small scale ($\lambda_0 \sim 5 \div 10 \hmp$).
 This situation
is clearly not the one  corresponding to the more
"optimistic" estimation of the homogeneity scale $\lambda_0$
\cite{slmp98,rees99,jmsl99}.
The other assumption usually made is that
galaxy positions are independent on their luminosity.
We have shown\cite{syl00}
 that, although such an assumption cannot
describe local morphological properties of 
galaxy distribution\cite{slp96}, it works rather well 
in the available galaxy redshift surveys.

The estimation of the the average density we are able
to make depends hence on two parameters. The first one 
is the 
 homogeneity scale $\lambda_0$ and the second 
is the Mass-to-Luminosity ratio. 
We can give an 
upper limit to $\Omega$ by taking the highest 
$({\cal M}/{\cal L})_c = 300h$ observed up to now \cite{bah99}
(in clusters of galaxies) to be universal across all the scales, 
and by considering
a lower limit for the homogeneity scale $\lambda_0= 50 \hmp$. 
We  compute the critical $({\cal M}/{\cal L})_{crit} $,
i.e. the Mass-to-Luminosity ratio needed to have $\Omega =1$.
As the others, also this 
parameter depends on the homogeneity scale $\lambda_0$.

\subsection{Average luminosity density  from galaxy catalogs}
Let
\be
\label{e1}
\langle \nu(r,L) \rangle dL d^3r  = 
\phi(L) \langle \Gamma(r)\rangle dL d^3r =
A r^{D-3} L^{\alpha} e^{-\frac{L}{L_*}} d^3r dL
\ee
be the average  number of galaxies  in the volume element
 $d^3r$ at distance $r$ from a observer
located on a galaxy,  
and with luminosity in the range $[L,L+dL]$. In Eq.\ref{e1}
we have used the fact that the galaxy luminosity function has 
been observed to have the so-called Schechter shape with parameters
$L_*$ (luminosity cut-off) and $\alpha$ (power law index)
which can be determined experimentally. The conditional average space density
 $\langle \Gamma(r)\rangle $ has a power 
 law behavior corresponding to a fractal
dimension $D$ (which eventually can be a function of scale, and 
hence can approach to $D=3$ at a scale $\lambda_0$).
Both the fractal dimension $D$ and the overall amplitude 
$A$ can be determined in redshift surveys. 
Hence $\langle \nu(r,L) \rangle$ is a function of 
four parameters: $L_*, \alpha, D, A$.
Moreover we note that by writing 
$\langle \nu(r,L) \rangle$ as a product 
of the space density and of the luminosity function
we have implicitly assumed that galaxy positions are independent
on galaxy luminosity.

 We would like to estimate the average luminosity density
 in a sphere of radius $R$ and volume $V(R)$ placed
 around a galaxy, and defined as 
\be
\label{e2}
\langle j(<R) \rangle = \frac{1}{V(R)} 
\int_0^R \int_{0}^{\infty} L\langle \nu(r,L)\rangle dL d^3r 
\equiv j(10) \left(\frac{R}{10 \hmp}\right)^{D-3}
\ee
which is $R$ dependent as long as the space density 
shows power law behavior (i.e. $D<3$).
By considering $M_*=19.53$ 
(i.e. $L_*=1.0 \cdot 10^{10} h^{-2} L_{\odot}$), $\alpha=-1.05$ \cite{lumfun} 
and
by estimating the prefactor $A$ (Eq.\ref{e1}) 
 and the fractal dimension ($D\approx 2$)  in galaxy redshift samples
we obtain\cite{syl00}
\be
\label{e7}
j(10) \approx 2 \cdot 10^{8} \;  h L_{\odot}/Mpc^3 \; .
\ee

We now estimate the density parameter in terms of the critical density,
\be
\label{e8}
\rho_{c} = 2.78 \cdot 10^{11} h^2 \; M_{\odot}/Mpc^3
\ee
where $M_{\odot}$ is the solar mass.
By considering the product of  the 
mass-to-luminosity ratio (in  solar  and  $h$ units) and the 
average luminosity density given by Eqs.\ref{e2}-\ref{e7}, we obtain
\be
\label{e9}
\Omega(\lambda_0) = (6 \pm 2) \cdot 10^{-4}  
\frac{\frac{{\cal M}} {{\cal L}}}{h}
  \left(\frac{\lambda_0}{10h^{-1}}\right)^{-1} \;  ,
\ee
where $\lambda_0$ is the scale where the crossover to
homogeneity occurs (it can also be  $\lambda_0 = \infty$, and 
in such a case $\Omega(\infty)=0$). Note that 
in view of the dependence of ${\cal M}/{\cal L}$ on $h$,
 Eq.\ref{e9}
does not depend on the Hubble's constant.

Let us now suppose that  ${\cal M}/{\cal L} \approx  10h$ as 
it has been derived by Faber and Gallengher \cite{fg79}.
We obtain
\be
\label{e11}
\Omega(\lambda_0) 
\approx  6  \cdot 10^{-3}\left(\frac{\lambda_0}{10h^{-1}}\right)^{-1} \; .
\ee
If galaxy distribution turns out to be homogeneous at 
$\lambda_0 \approx 10 \hmp$ then $\Omega \approx 6  \cdot 10^{-3}$ as
it is obtained in the standard treatment \cite{pee93}.
If, instead, the crossover to homogeneity lies at $100 \hmp$,
we obtain $\Omega \approx  6 \cdot 10^{-4}$.

From Eq.\ref{e9} and if galaxy distribution turns out 
to be homogeneous at $\lambda_0$,
we obtain that the critical Mass-to-Luminosity ratio
(such that $\Omega(\lambda_0)=1$) is given by
\be
\label{cr1}
\left(\frac{{\cal M}}{{\cal L}}\right)_{crit} \approx  1600h  
\left(\frac{\lambda_0}{10h^{-1}}\right) \;, 
\ee
so that if $\lambda_0=10 \hmp$ one obtains 
$({\cal M}/{\cal L})_{crit} \approx  1600h$
(which is again consistent with the usual
 adopted value \cite{pee93})
while if $\lambda_0=100 \hmp$ $({\cal M}/{\cal L})_{crit} \approx  16000h$,
which is about two orders of magnitude larger
than the highest $ {\cal M}/{\cal L}$ observed in clusterised
objects.
For an intermediate value of $\lambda_0= 50 \hmp$
one obtains $({\cal M}/{\cal L})_{crit} \approx  8000h$.

For what concerns the analysis of 
galaxy clusters it is often used a value $({\cal M}/{\cal L})_{c} \sim 300h$
\cite{galclu,bah99}
which, by using Eq.\ref{e9}, gives 
\be
\label{c1}
\Omega(\lambda_0) \approx  
2 \cdot 10^{-1}\left(\frac{\lambda_0}{10h^{-1}}\right)^{-1} \; .
\ee
Such an estimation is based on the fact that
the Mass-to-Luminosity ratio found in clusters 
is representative of all the field galaxies. 
This is a very strong assumption:
this means that 
$({\cal M}/{\cal L})_{g} = 
({\cal M}/{\cal L})_{c}$ which is not supported
by any observation.
The usually adopted value of $\Omega =0.2$ \cite{bah99}
can be derived from Eq.\ref{c1} by assuming $\lambda_0= 10 \hmp$.
In the case the crossover to homogeneity
occurs at $\lambda_0=100 \hmp$ we have that
$\Omega(\lambda_0) = 0.02$ if we consider 
$({\cal M}/{\cal L})_{c} \sim 300h$ to be "universal".

Under the assumptions:

{\it (i)} galaxies are homogeneously distributed
at scales larger than $\lambda_0 \approx 5 \hmp$, 

{\it (ii)} the ${\cal M}/{\cal L}$
of galaxies is the same of clusters,
that is galaxies should contain a factor $\sim 30$ more
dark matter than what it is observed
with the study of the rotation curves \cite{fg79},

{\it (iii)} Galaxy positions are independent of galaxy luminosity:
such a assumption is not strictly valid, but it has been 
tested\cite{syl00} to hold rather well in the available samples,

we  get the following  upper limit to $\Omega$.
If we assume that ${\cal M}/{\cal L} = 300 h$
across all the scales, and that 
$\lambda_0 \approx 50 \hmp$,
which we consider to be a lower limit for the homogeneity scale,
we get from Eq.\ref{c1}
\be
\label{ul1}
\Omega(50 Mpc/h, 300 h M_{\odot}/L_{\odot}) \le 0.04 \;.
\ee

The direct estimation   with galaxies (Eqs.\ref{e9}-\ref{e11})
gives lower a value, the reason being 
the strong assumption of taking   ${\cal M}/{\cal L} = 300 h$
as representative of field galaxies (see Fig.\ref{figomega}).
\bef 
\epsfxsize 8cm  
\epsfysize 8cm  
\centerline{\epsfbox{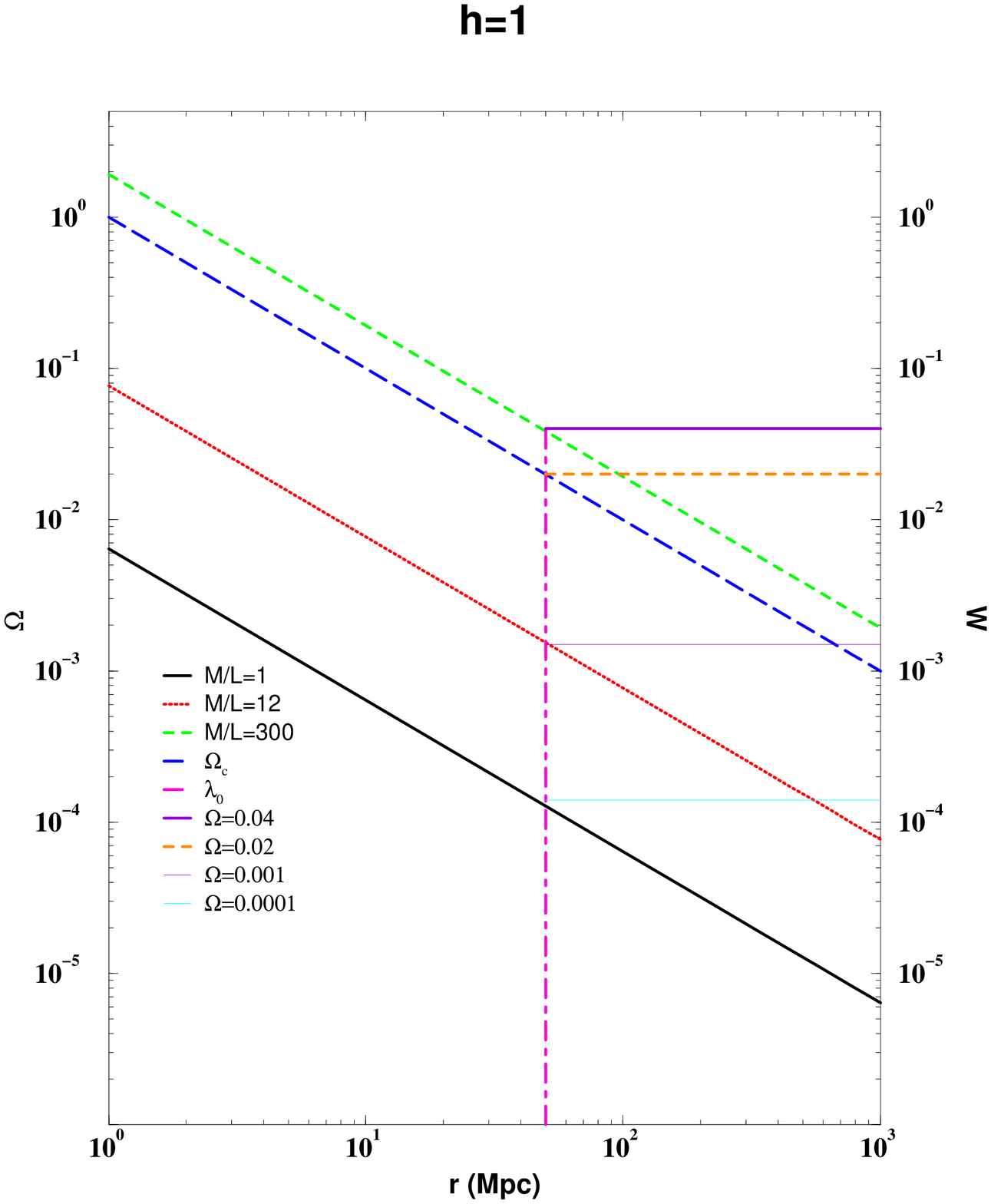}}  
\caption{\label{figomega}
We show the behavior of $\Omega(r,{\cal M}/{\cal L})$
and $\Omega_c(r)$ (direct estimation from the Mass and number density
of galaxy clusters),
in the case $h=1$,
for three different values of ${\cal M}/{\cal L}=1, 12, 300$.
If $\lambda_0 =50 \hmp$ we get an upper limit
$\Omega =0.04$. In the figure the value of the 
Hubble's constant is set $h=1$. (From Sylos Labini, 2000)
}
\eef

\subsection{Homogeneity scale and primordial nucleosynthesis constraints}

Let us now discuss the previous 
results in relation with the nucleosynthesis
constrain $\Omega_b^{BBN}  = 0.015h^{-2}$, deriving in such a
way an upper limit for $\lambda_0$ which is consistent
with such a scenario.
We may see\cite{syl00}  
 that if $\lambda_0 \approx 100 h^2 Mpc$ then
$\Omega(\lambda_0,({\cal M}/{\cal L})_c) \approx \Omega_b^{BBN}$.
Such a fact has two important implications.

First of all one does not need the presence 
of {\it non baryonic dark matter} to reconcile 
local observations of matter contained in galaxies
and galaxy clusters ($\Omega_{local}$)
with the primordial nucleosynthesis
constraint. This is an important point as the 
existence of {\it non baryonic dark matter}
has been inferred also, but not only,
by the discrepancy $\Omega_{local}/\Omega_b^{BBN}$ \cite{bah99}, 
which, we show, is not observed if $\lambda_0 \ge 100 h^2 Mpc$.
There is still place, in this picture,
for a uniform background of {\it non-baryonic dark matter},
which has completely different clustering properties 
from the ones of visible matter.

The second point concerns the baryon fraction  in clusters.
According to the usual root, in  order to be consistent 
with $\Omega_b^{BBN}$,
one has to force $\Omega =0.1 \div 0.3$ from clusters 
analysis\cite{bah99}. Here we can argue as follows.
By assuming a very large ${\cal M}/{\cal L} \sim 300 h$,
and that it is universal across all scales,
and by assuming that all dark matter  is made
of baryons, if the crossover to homogeneity occurs at scales larger 
than $\sim 150 h^2 Mpc$ then there  is not enough baryonic
matter to satisfy the nucleosynthesis constraint. 
The situation gets clearly worst if one takes into account
that not all the mass in clusters is baryonic 
or that ${\cal M}/{\cal L} < 300 h$, i.e. it is not considered
to be universal across all scales,  or that
$\lambda_0 \gg 100 h^2 Mpc$. 
There are then two different   possibilities:
the first is to study the case of a non homogeneous primordial
nucleosynthesis which could lower the limit
on $\Omega_b^{BBN}$, given the observed
abundances of light elements, and the second
would be to have a more uniform background of baryonic dark matter.
This seems rather unlikely because it would have very different
clustering properties with respect to visible mass
and it is very difficult to find a 
dynamical explanations for such a 
segregation.

\subsection{ Where does linear approximation hold ?}
In the standard picture, 
the properties of dark matter on 
cosmological relevant scales $r > 5 \hmp$,
are inferred from the observed  
properties of visible matter and radiation.
Now one studies change in these properties,
i.e. the presence of fractal correlations, 
and in this respect they will have consequences 
on dark matter too\cite{dsl98,gsl00}.
 For example, the determination of the 
mass density including dark matter has been performed
on the basis of the linear theory \cite{sw94}.
Here the problem is: beyond which
scale can linear theory  be considered
as a useful approximation ?

In other words, the dynamical estimates use gravitational effects,
of departure from a strictly homogeneous distribution
on the motion of objects such as galaxies
considered as a test particle.
A completely different situation occurs if $\lambda_0$ is larger 
than the scales at which linear approximation is usually adopted.
For example, methods based on the Cosmic Virial Theorem \cite{pee80},
distortions in redshift surveys \cite{sw94}, local group dynamics \cite{pee93},
the "reconstructed" peculiar velocity
field from the density field (e.g POTENT-like methods\cite{sw94}),
are clearly not useful at scales $r \ll \lambda_0$.
Up to now it is implicitly
 assumed $\lambda_0 \approx 5 \div 10 \hmp$
and all these methods are considered to be valid on larger scales. 
If, for example $\lambda_0 \sim 50 \hmp$
then it is not possible to interpret   peculiar velocities
in the range $1 \div 30 \hmp$ through the linear 
approximation (as it is usually done\cite{sw94}),
 unless there is a  
background of dark matter which is uniform beyond
a certain small scale. In such a situation however
the estimates would be different \cite{bar98}
 from the usual ones \cite{sw94}.

Let us review some simple relations between 
the conditional average density $\Gamma(r)$ and the usual
correlation function $\xi(r)$.
If $\Gamma(r)$ has a power law behavior up to a scale $\lambda_0$
and then it presents a crossover to homogeneity,
it is simple to show that (in the case of $D=2$)
the length scale at which $\xi(r_0)=1$ is 
of the order of
\be
\label{o1}
r_0 \approx \frac{\lambda_0}{3}
\ee
where the exact relation between $r_0$ and $\lambda_0$
depends on the details of the crossover \cite{gsl00}.
However Eq.\ref{o1} gives a reasonable order of magnitude
in the general case.
In such a situation it is also simple to
compute $\sigma(r,\lambda_0) 
= \langle N(r) - \langle N \rangle / \langle N \rangle \rangle$,
i.e the amplitude of the fluctuation with respect
to the average at the scale $r \ll \lambda_0$ 
\cite{slmp98}
\be
\label{o2}
\sigma(r,\lambda_0) \approx \frac{\lambda_0}{2r} \; .
\ee
It has been found in various nearby surveys that $\sigma(8 \hmp) =1$.
However if the crossover to homogeneity occurs at $\lambda_0$
we have that
\be
\label{o3}
\sigma(8 \hmp,\lambda_0) \approx \frac{\lambda_0}{16}
\ee
For example if $\lambda_0 \approx 15 \hmp$ then $r_0\approx 5 \hmp$
and $\sigma(8 \hmp,15 \hmp) = 1$;
otherwise if $\lambda_0 \approx 50  \hmp$ then $r_0\approx 16 \hmp$
and $\sigma(8 \hmp,15 \hmp) = 3$.
 
Linear approximation holds in the linear regime,
when the amplitude of the density contrast is small, 
i.e. for $\sigma \ll 1$.
We have that $\sigma = 1 $ at a scale $r_{\sigma=1}(\lambda_0)$ 
\be
\label{o4}
r_{\sigma=1}(\lambda_0)\  \approx \frac{\lambda_0}{2}
\ee
For instance, if the crossover to homogeneity occurs at 
$\lambda_0 = 50 \hmp$ one has that $\sigma(\lambda_0)=1$
at $r_{\sigma=1}= 25 \hmp$. In such a situation all the
estimations of the density parameter at smaller scales 
based on the linear approximation\cite{sw94}, and hence based
on the untested assumption of linearity, are not correct.

We have studied in detail\cite{gslp99} the gravitational force distribution
in a fractal structure.
Its behaviour can be 
understood as the sum of two parts, a local or `nearest neighbours'
piece due to the smallest cluster (characterised by the lower
cut-off $\Lambda$ in the fractal) and a component coming from the 
mass in other clusters. The latter is bounded above by the
scalar sum of the forces 
\begin{equation}
\langle |\vec{F}| \rangle \leq 
\lim_{L\rightarrow \infty} 
\int_{\Lambda}^L \frac{G\rho_m(r)}{r^2} 4\pi r^2 dr \sim  L^{D-2}
\end{equation}
so that for $D<2$ it is convergent, while for $D>2$
it may diverge. If there is a divergence, it is due 
to the presence of angular fluctuations at large scales,
described by the three-point correlation properties of
the fractal. For the difference in the force between two points
the local contribution will be irrelevant well beyond the scale
$\Lambda$, while it is easy to see that the `far-away' contribution
will now converge as $L^{D-3}$, and its being non-zero 
is a result of the absence of perfect spherical 
symmetry. We have then applied such a result to the case
of an open universe\cite{jampsl00} in order to compute
the expected deviations from a pure linear Hubble flow.

\section{Discussion and Conclusions}

We now present a short discussion about the perspective of our work:
the cosmological implications of the fractal behavior of
visible matter crucially depend on the crossover scale $\lambda_0$,
but, no matter what is the actual value of such a scale,
we  have some important consequences from 
the theoretical point of view. We may identify three
different scenarios.

\begin{itemize}

\item (1) The fractal extends only up to $\sim 30\div 100 \hmp$.
This is the minimal concept which begins to be 
absorbed in the literature\cite{rees99} but, 
sometimes, without considering its real consequences.
The standard approach to galaxy distribution
has identified very small "correlation lengths",
namely $5 \hmp$ for galaxies and $25 \hmp$ for clusters.
These numbers (which were supposed to know with high 
precision\cite{dav97}) are anyhow inconsistent 
with the fractal extending a factor $2\div 4$ more.
We have shown that this inconsistency is conceptual and not due to 
incomplete data or week statistics. Hence, in this  
hypothesis one has to abandon all the concepts
related to these length scales. These are: 

 (i) 
The estimate of the matter density in clusterised objects
(visible + dark), which has been claimed to be 
$\Omega \approx 0.2 \div 0.3$, decreases 
by one order of magnitude or more. 

(ii) The normalization of N-body simulations 
is usually performed to some length-scale
or amplitude of fluctuations, which are related to
$5 \hmp$ and $25 \hmp$.

(iii) Concepts like the galaxy-cluster mismatch 
and the related luminosity bias, as well as 
the understanding of the clustering via
the bias parameter $b$ (i.e. linear or non-linear - "stochastic bias" -
amplification of $\xi(r)$) loose any physical meaning.

(iv) The interpretation of the velocity field is 
also based on the linear approximation which
cannot certainly hold at scales smaller than $30\div50 \hmp$.

(v) The reconstruction of the three dimensional properties
 from the  angular data suffers of the same untested
 assumption of homogeneity.

In summary major modifications are necessary 
for the origin and dynamics of large scale structures
and for the role  of dark matter. However the structures
may still formed via gravitational instability,
in the sense that 
they are not necessarily primeval\cite{jampsl00}.

\item (2) The fractal extends up to $300 \div 500 \hmp$.
In this case the standard picture of
gravitationally induced structures after 
the electromagnetic decoupling is untenable. There
is no time to create such large scale correlated structures
via gravitational instability, starting from Gaussian
initial conditions. More string consequences 
are clearly important for what concerns 
the amount of matter in clusterised objects.

\item (3) The fractal extends up to $\gtapprox 1000 \hmp$,
and homogeneity does not exist, at least 
for what concerns galaxies. In this extreme case
a new picture for the global metric\cite{jampsl00} 
is then necessary.

\end{itemize}

For some questions 
the fractal structure leads to a radically new perspective  
and this is hard to accept. But it is based on the best data and  
analyses available. It is neither a conjecture nor a model,  
it is a fact. 
The theoretical problem is that  
there is no dynamical theory to explain 
how such a fractal Universe could have arisen from the pretty 
smooth initial state we know existed in the big bang.  
However this is a  different  
question. The fact that something can be hard to explain theoretically has 
nothing to do with whether it is true or not. Facing a hard problem is far 
more interesting than hiding it under the rug by an inconsistent  procedure.
For example some interesting attempts to understand why 
gravitational clustering generates scale-invariant structures have been
recently proposed by de Vega et al\cite{devega1,devega2,devega3}.
Indeed this will be the key point to understand in the future,  
but first  we should agree on how these new 3d data should be analyzed. 
In addition, the eventual crossover to homogeneity has also to be found with 
our approach. 
If for example homogeneity would really be found say at  
$  \sim100 \hmp$, then clearly 
all our criticism to the previous methods
and results  still holds fully. In summary the 
standard method cannot be used neither to disprove homogeneity, nor to 
prove it. One has simply to change methods.

\section*{Acknowledgements} 
 
We warmly thank   
R. Durrer, A. Gabrielli, 
M. Joyce and M. Montuori  
with whom various parts of this work have 
been done in  fruitful collaborations. 
FSL is grateful to Pedro Ferreira for very useful 
comments and discussions. 
We thank Prof. N. Sanchez 
for the organization of this very interesting School.  
This work has been partially supported by the  
EC TMR Network  "Fractal structures and   
self-organization"   
\mbox{ERBFMRXCT980183} and by the Swiss NSF.

\end{document}